 \documentclass[final,3p,times,twocolumn]{elsarticle}

\usepackage{amssymb}
\usepackage{amsmath}

\begin{document}

\begin{frontmatter}



\title{Separation of atomic and molecular ions by ion mobility with an RF carpet}

\author[1]{Ivan Miskun\corref{cor1}}
\author[1,2]{Timo Dickel}
\author[1,2]{Samuel Ayet San Andr\'{e}s}
\author[1]{Julian Bergmann}
\author[3]{Paul Constantin}
\author[1]{Jens Ebert}
\author[1,2]{Hans Geissel}
\author[1]{Florian Greiner}
\author[2]{Emma Haettner}
\author[1]{Christine Hornung}
\author[1]{Wayne Lippert}
\author[4,5]{Israel Mardor}
\author[6]{Iain Moore}
\author[1,2]{Wolfgang R.~Pla\ss}
\author[2]{Sivaji Purushothaman}
\author[1]{Ann-Kathrin~Rink}
\author[1,7]{Moritz P.~Reiter}
\author[1,2]{Christoph~Scheidenberger}
\author[2]{Helmut Weick}

\cortext[cor1]{Corresponding author: \mbox{ivan.miskun@exp2.physik.uni-giessen.de}}

\address[1]{II.~Physikalisches Institut, Justus-Liebig-Universit\"at Gie\ss en, 35392 Gie\ss en, Germany}
\address[2]{GSI Helmholzzentrum f\"ur Schwerionenforschung GmbH, 64291 Darmstadt, Germany}
\address[3]{IFIN-HH/ELI-NP, 077125, M\u{a}gurele - Bucharest, Romania}
\address[4]{Tel Aviv University, 6997801 Tel Aviv, Israel} 
\address[5]{Soreq Nuclear Research Center, 81800 Yavne, Israel}
\address[6]{University of Jyv\"askyl\"a, 40014 Jyv\"askyl\"a, Finland}
\address[7]{University of Edinburgh, EH8 9AB Edinburgh, United Kingdom}

\begin{abstract}
	Gas-filled stopping cells are used at accelerator laboratories for the thermalization of high-energy radioactive ion beams. Common challenges of many stopping cells are a high molecular background of extracted ions and limitations of extraction efficiency due to space-charge effects. At the FRS Ion Catcher at GSI, a new technique for removal of ionized molecules prior to their extraction out of the stopping cell has been developed. This technique utilizes the RF carpet for the separation of atomic ions from molecular contaminant ions through their difference in ion mobility. Results from the successful implementation and test during an experiment with a 600~MeV/u $^{124}$Xe primary beam are presented. Suppression of molecular contaminants by three orders of magnitude has been demonstrated. Essentially background-free measurement conditions with less than $1~\%$ of background events within a mass-to-charge range of 25 u/e have been achieved. The technique can also be used to reduce the space-charge effects at the extraction nozzle and in the downstream beamline, thus ensuring high efficiency of ion transport and highly-accurate measurements under space-charge-free conditions.
\end{abstract}


\end{frontmatter}


\section{Introduction}
\label{Introduction}

Gas-filled stopping cells are widely used in many in-flight accelerator facilities to slow down high-energy radioactive ion beams (RIBs) and to enable high-accuracy measurements with low-energy ions \cite{Wada2013}. Incoming ions are thermalized in a noble gas, extracted and delivered to the measuring device. For efficient thermalization of ion beams produced at relativistic energies, areal densities of the buffer gas of at least a few mg/cm$^2$ are required. In order to provide a fast extraction and, therefore, give access to short-lived nuclei, the ions are extracted by a combination of DC and RF electric fields and a gas flow \cite{Wada2003,Savar2003,Weiss2005,Neuma2006a,Ranja2011}. RF fields are created by so-called RF carpets and funnels, devices with fine electrode structures that repel ions, preventing them from being lost and guiding them towards an extraction orifice.

An important limitation of the extraction efficiency, encountered by high-density stopping cells, is caused by impurities of the buffer gas. During slowing down and thermalization in the buffer gas, each incoming ion generates about $2.5\times10^4$ ion-electron pairs per MeV of deposited energy in a helium buffer gas. The ionized buffer gas atoms may undergo charge-exchange reactions with atoms of impurities contained in the buffer gas. Once ionized, the contaminants can form adducts and molecules with mass-to-charge ratios over a very broad range and are extracted out of the stopping cell together with the ion of interest (IOI) \cite{Kudry2001}. Adducts are loosely-bound and can be fragmented after their extraction with dissociation techniques, such as collision-induced dissociation (CID) \cite{Mcluc1992,Schur2006}. Strongly-bound molecules are harder to remove. They lead to potentially harmful space charge in the downstream transport beamline and high-precision experimental devices and, in some cases, might even deteriorate the accuracy of the measurement.

In addition, the extraction efficiency of the stopping cell can also be limited by space-charge effects occurring inside of the cell \cite{Morri2007,Moore2008}. Electrons, due to their high mobility, are removed with electric fields out of the stopping volume very quickly. In contrast, the positively-charged ions of the buffer gas and its contaminants have velocities similar to the one of the stopped IOI. At high rates of the incoming beam, this can lead to a build-up of a positive charge inside the stopping volume of the stopping cell. A significant space charge can distort the electric fields and cause additional ion losses in the bulk of the cell \cite{Takam2005,Reite2016} or in the extraction orifice \cite{Sumit2020}. The space-charge effects in the extraction orifice are not universally observed in all gas-filled stopping cells, however, in some cases, they can cause drastic efficiency losses. This limitation is currently one of the main bottlenecks for the production of cooled and re-accelerated RIBs of high intensities.

In this work, a novel separation technique that allows molecular suppression prior to ion extraction out of the stopping cell is presented. This technique utilizes the dependence of the effective repelling field of the RF carpet on ion mobility. It has been used to avoid space charge at the extraction nozzle of a stopping cell and for suppression of molecules, in particular of those that are too strongly-bound to be dissociated by the CID in the RFQ beamline.

\section{Experimental Method}
Ion mobility spectrometry is commonly used in analytical chemistry and has many applications \cite{Eicem2013}. There, ion funnels are often employed for radial focusing, accumulation, and bunching of ions \cite{Kelly2010}. In this work, we propose to use the RF funnel or carpet itself for the separation by ion mobility.

The effective repelling field $ E_{\mathrm{eff}} $ of the RF carpet depends on different parameters \cite{Tolma1997, Wada2003} and can be described from the pseudopotential approach by
\begin{equation}
\label{Eq_1_19}
E_{\mathrm{eff}}=2 V_{\mathrm{RF}}^{2}~\frac{1}{r_{0}^{3}} \left (\frac{r}{r_{0}} \right )\left (K_{0}\frac{T}{T_0}\frac{p_0}{p}\right )^{2} \frac{m}{q}, \end{equation}
where $ m $ and $ q $ are the mass and the charge of the ions, $ V_{\mathrm{RF}} $ is the amplitude of the applied RF voltage, $ r_0 $ is the half of the electrode pitch size (the distance between centers of neighboring electrodes), $ K_0 $ is the reduced ion mobility, $T$ is the temperature ($ T_0=273.15 $~K), and $p$ is the pressure of the buffer gas ($ p_0=1.013 $~bar). For fixed operating parameters, ions with different mass-to-charge ratios and mobility values experience different effective repelling field. From Eq.\ (\ref{Eq_1_19}), it is possible to deduce the dependence of the relative change in the effective repelling field of the RF carpet $\Delta E_{\mathrm{eff}}/E_{\mathrm{eff}}$ on relative changes in the mass-to-charge ratio and the mobility of the ion:
\begin{equation}
\label{Eq_3_1}
\frac{\Delta E_{\mathrm{eff}}}{E_{\mathrm{eff}}}=\frac{\Delta (m/q)}{(m/q)}+\frac{2\Delta K_0}{K_0}.
\end{equation}

For ions with a value of $K_0^2\times(m/q)$ lower than a certain threshold value, the effective repelling field is not strong enough to compensate the DC field pushing the ions towards the carpet. These ions are not repelled by the RF carpet and are lost on the surface of the electrodes. This is of great importance as it helps to prevent the extraction of low-mass ionization products out of the stopping cell, in particular the ionized buffer gas. The dependence of $E_{\mathrm{eff}}$ on the reduced ion mobility $ K_0 $ makes it possible to separate the IOI and molecular contaminants with the same mass-to-charge ratio by their respective ion mobilities with the RF carpet.

The separation by ion mobility at the RF carpet works in the following way. In a buffer gas, molecules and adducts have lower mobility value than atomic ions of the same mass-to-charge ratio, due to their larger size. This means that, according to Eq.~(\ref{Eq_1_19}), for the same operating conditions, the molecular ions have a weaker effective repelling field than the isobaric atomic ions. Thus, with the proper choice of the RF amplitude, it is possible to find a regime whereby the atomic ions are repelled and extracted out of the stopping cell, and the molecular ions are lost on the electrodes of the RF carpet. Furthermore, in helium gas at cryogenic temperatures of 70-80~K, atomic positively-charged ions over a broad mass-to-charge range have similar reduced mobility values within a typical range of 15 to 21~cm$ ^{2} $/(Vs) \cite{Viehl2012,Visen2020}. The major charge carrier in helium buffer gas at these temperatures, which is He$^+_3$, also has comparable ion mobility of about 18~cm$ ^{2} $/(Vs) \cite{Patte1970}. Therefore, it is possible to find operating conditions that allow the transport of atomic ions in the mass-to-charge region of interest while the molecular contaminants are suppressed. As can be seen from Eq.~(\ref{Eq_3_1}), the effective repelling field is twice as sensitive to the change in the ion mobility than to the change in the mass-to-charge ratio, which allows molecular suppression without significant losses for atomic ions over a broad mass-to-charge range.

\section{Experimental Setup}
\begin{figure}[]
\centering
\includegraphics[width=0.45\textwidth]{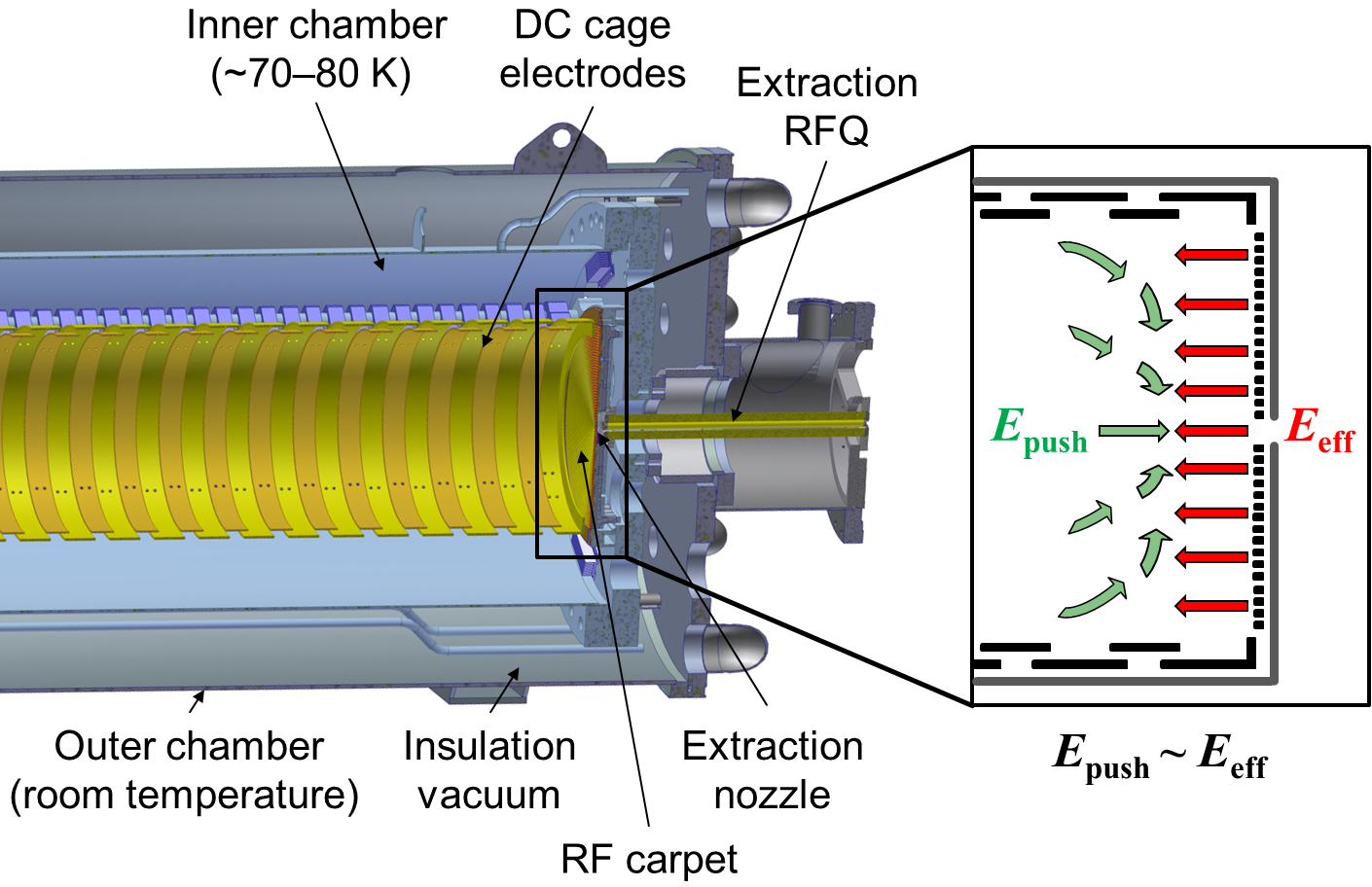}
\caption{Left: schematic figure of the part of the cryogenic stopping cell (CSC) in a sectional view. Right: schematic representation of the pushing and focusing DC fields and the counteracting effective repelling field of the RF carpet.}
\label{Fig_1}
\end{figure}

The FRS Ion Catcher (FRS-IC) \cite{Plass2013b,Plass2019} is a test facility for the Low-Energy Branch of the Super-FRS \cite{Geiss2003,Winfi2013} at FAIR. It is installed at the end of the symmetric branch of the fragment separator FRS \cite{Geiss1992} at GSI and consists of three major parts: a gas-filled cryogenic stopping cell (CSC) \cite{Ranja2011,Reite2015,Ranja2015}, an RFQ beamline \cite{Reite2015} and a multiple-reflection time-of-flight mass spectrometer (MR-TOF-MS) \cite{Plass2008,Dicke2015a}. The exotic nuclei are produced, separated in-flight and range bunched at relativistic energies of up to 1 GeV/u in the FRS, slowed down in degraders, and thermalized in the CSC. The thermalized ions are guided to the exit side of the stopping cell by a DC electric field. At the exit side, a PCB-based RF carpet with concentric electrodes with a density of 4~electrodes per millimeter is installed. The RF carpet repels and guides the ions towards the extraction nozzle with a combination of RF and radial DC electric fields. This part of the CSC, as well as the electric fields in the vicinity of the RF carpet, are shown schematically in Fig.~\ref{Fig_1}. Once the ions reach the nozzle, they are extracted by the gas flow into the RFQ beamline and transported to the MR-TOF-MS, where high-resolution mass measurements are performed.

For the CSC, many different technological steps were taken to ensure the high cleanliness of the buffer gas. To reduce the amount of residual gas inside the CSC, the CSC is baked at a temperature of about 400~K and constantly pumped by a turbomolecular pump for a few days before the cool-down. This, together with cryogenic operation temperatures of 70-80 K, helps in removing most of the contaminants \cite{Ranja2015}. The stopping cell itself works as a cryopump, where residual contaminants freeze out on the walls of the device. In addition, before being supplied into the CSC, the helium buffer gas is purified by an LN$_2$ cold trap and by a getter (MicroTorr, SAES Pure Gas). In combination, all these steps result in the high cleanliness of the CSC, which is required for ion survival and contaminant suppression.

Nevertheless, some contaminants still enter the CSC in trace amounts together with the helium buffer gas. Mostly, these are noble gases present in helium gas cylinders as residuals. They are very hard to filter out completely. Atoms of noble gases contained in the buffer gas do not harm the ion survival, but can be ionized and extracted when the incoming ion beam is thermalized in the stopping cell. It has been seen that the rates of extracted ions of different noble gases change from one gas cylinder to another.

The separation by ion mobility was investigated in an online experiment at the FRS-IC. The $ ^{124} $Xe primary beam was stopped in the CSC and produced ionization of the buffer gas and its contaminants. Extracted ionization products were measured with the MR-TOF-MS in the broadband time-focus shift (TFS) mode \cite{Dicke2015a} at a mass resolving power of about 1000. The CSC was filled with 75~mbar of helium gas at a temperature of 74~K, which corresponds to an areal density of 5.1~mg/cm$^2$. A DC push field of 14~V/cm was applied. The ions spent 0~to~170~ms on the carpet surface before their extraction due to a problem with the electronics of the RF carpet, which, nevertheless, did not affect the extraction efficiency \cite{Misku2019a,Misku2019b}. The RF carpet was operated at a frequency of $ 5.92 $~MHz. In all measurements, the extraction RFQ was operated as a mass filter, transmitting a mass-to-charge window from $ 70 $ u/e to $ 95 $ u/e. The isolation-dissociation-isolation technique (IDI) \cite{Grein2019} was used as a primary method for the suppression of molecular background. It was implemented by using the RFQ mass-filter as the first isolation step, CID in the RFQ beamline with a voltage step of 50 V, and a low-mass cut-off of the first RFQ of the MR-TOF-MS as the second isolation step. In the mass-to-charge region of the $ ^{124} $Xe$^+$ ions extracted out of the CSC, all molecular contaminants could be efficiently removed from the measured spectrum by the IDI technique only \cite{Grein2019}.

\section{Results}
\begin{figure}[]
\centering
\includegraphics[width=0.45\textwidth]{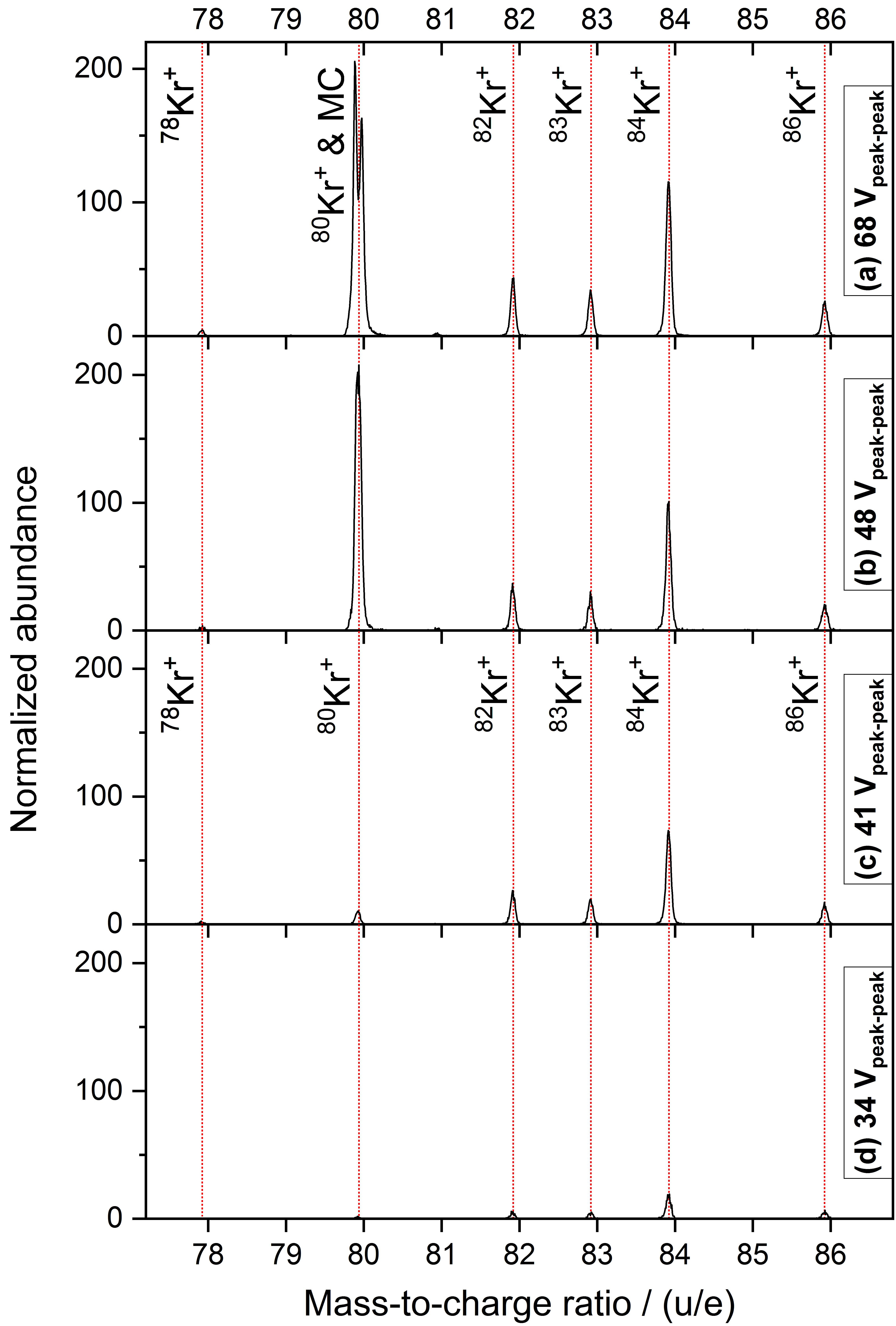}
\caption{Mass spectra of ionization products extracted out of the CSC measured with the MR-TOF-MS in the TFS mode at a mass resolving power of $\sim 1000$. Stable isotopes of krypton and molecular contaminant (MC) are present. IDI is applied. All spectra are normalized to the number of spills and intensity of the incoming ion beam. a) RF carpet is at 68~V$ _{\mathrm{peak-peak}} $, abundant molecular contaminant overlapping with $^{80}$Kr$^+$ is not removed by IDI; b) RF carpet is at 48~V$ _{\mathrm{peak-peak}} $, no change in spectrum composition despite $ 30~\% $ lower RF repelling voltage, the molecular contaminant is partly suppressed but not removed; c) RF carpet is at 41~V$ _{\mathrm{peak-peak}} $, the molecule at 80~u/e is removed, the abundance of the peak at $^{80}$Kr$^+$ corresponds to the natural abundance ratio of stable Kr isotopes; d) RF carpet is at 34~V$ _{\mathrm{peak-peak}} $, Kr isotopes are suppressed.}
\label{Fig_2}
\end{figure}

The rate of the incoming ion beam was about 20,000~ions/s. The stopping efficiency of the CSC in this experiment was about $25~\%$, so $\sim5,000$~$^{124}$Xe ions per second were thermalized in the effective volume of the CSC. ATIMA calculations \cite{Weick2018} estimate an average energy deposition by a single incoming ion of 190 MeV, which corresponds to about $9\times10^{10}$~He$^+$-electron pairs per second generated in the CSC. In Figure~\ref{Fig_2}a, the measured mass-to-charge spectrum of lower-mass ionization products is shown. IDI was applied, the RF carpet was operated at $ 68 $ V$ _{\mathrm{peak-peak}} $. Most of the peaks in the spectrum correspond to the ionized stable isotopes of krypton, a residual contaminant of the helium gas from a gas cylinder. Because for several mass lines, the rate of detected ions was larger than 100 ions/s, dead-time effects occurred in the data acquisition system of the MR-TOF-MS for these mass lines. As a result, the measured relative abundances of the krypton isotopes do not precisely correspond to the literature values. The measured counts in each spectrum were normalized to the number of spills and intensity of the incoming ion beam. As can be seen from the spectrum, one isotope of krypton, $ ^{80} $Kr, overlaps with a high-abundant molecular contaminant. The dip in the middle of the peak does not represent a mass separation of two species but is caused by severe dead-time effects. With the IDI method, only about half of this molecular contaminant could be broken up. Besides the molecule at $ 80 $ u/e, there is no significant molecular background. For every few projectile ions hitting the CSC, only one Kr ion is produced. The measurement was done with a single helium gas cylinder, so the production rate of the Kr ions can be assumed to be constant. These results demonstrate the good cleanliness of the CSC.

Then, the RF voltage of the RF carpet was lowered from $ 68 $ V$ _{\mathrm{peak-peak}} $ to $ 48 $ V$ _{\mathrm{peak-peak}} $, which did not bring a significant difference to the composition of the spectrum, however, led to the partial suppression of the molecular contaminant, as can be seen from the reduction of dead-time effects in the peak at 80~u/e (Fig.~\ref{Fig_2}b). Afterwards, the RF voltage was further decreased down to $ 41 $ V$ _{\mathrm{peak-peak}} $. The spectrum for this setting is shown in Fig.~\ref{Fig_2}c. The molecular contaminant at the mass line $ 80 $ u/e was removed, leaving the $ ^{80} $Kr isotope peak, which fits its natural abundance. This behavior corresponds to the RF carpet operation regime when the effective repelling field is strong enough to repel the atomic ions, but not the molecular ones with lower ion mobility. The transition from the transmission of the molecular contaminant to its significant suppression is rather sharp for a small change in the RF voltage. This indicates that the separation by ion mobility is a very well-defined suppression method.

Further decrease of the repelling RF voltage to $ 34 $~V$ _{\mathrm{peak-peak}} $ led to the suppression of Kr isotopes, as shown in Fig.~\ref{Fig_2}d. This effect occurs due to the mass dependence of the effective repelling field of the RF carpet. The ions of lower masses feel a weaker effective repelling field.

\begin{figure}[]
\centering
\includegraphics[width=0.45\textwidth]{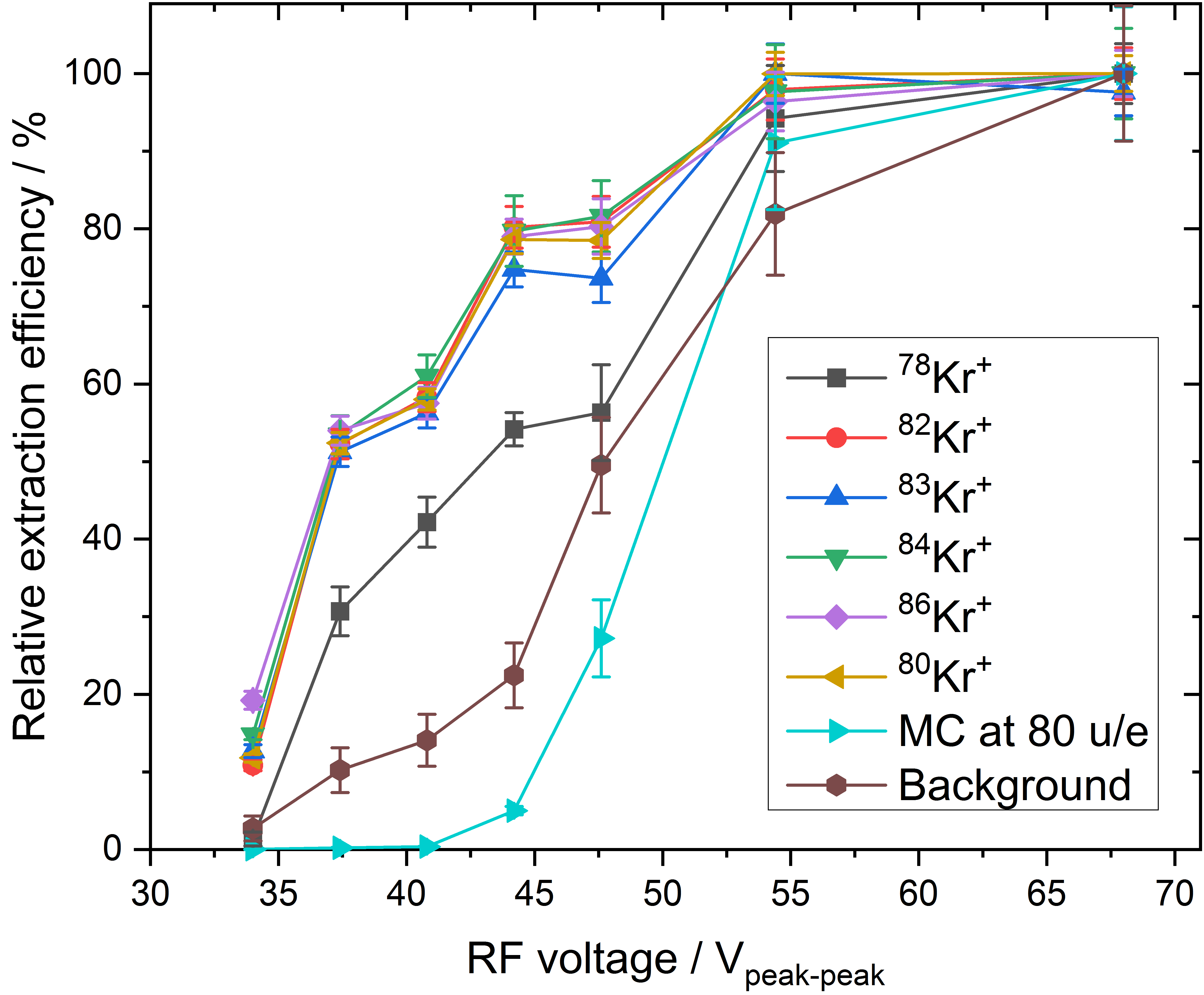}
\caption{Relative extraction efficiencies of stable Kr isotopes, molecular contaminant (MC) with $m/q$=80 u/e, and overall molecular background at different RF voltages applied to the RF carpet. Efficiencies of $^{80}$Kr$^+$ were estimated from the measured counts of $^{82}$Kr$^+$ and $^{83}$Kr$^+$ and their natural abundances. Efficiencies of MC were calculated from the total measured counts in 80 u/e peak and estimated $^{80}$Kr$^+$ efficiencies. A dramatic efficiency drop for the molecular contaminant is caused by its lower ion mobility value.}
\label{Fig_3}
\end{figure}

The relative extraction efficiencies of the measured ions are plotted in Fig.~\ref{Fig_3}. The measured efficiencies for Kr isotopes with masses 78, 82, 83, 84, and 86 u were normalized independently to their highest count rate detected within the RF voltage scan. The measured counts were corrected for the dead-time effects and the fluctuations in the beam intensity. The efficiency of the $ ^{80} $Kr isotope was estimated from the measured efficiencies of $ ^{82} $Kr and $ ^{83} $Kr and their natural abundance ratio. The efficiency of the molecular contaminant at the mass-to-charge line $ 80 $ u/e was calculated from the measured efficiency of all ions with $ 80 $ u/e and the estimated efficiency of the $ ^{80} $Kr isotope. Plotted error bars reflect the statistical uncertainties and uncertainties introduced by dead-time effects and changes in beam intensity. RF voltages could be measured with an accuracy of $\sim5~\%$. As can be seen from the plot, most stable Kr isotopes follow the same trend in extraction efficiency, which decreases with smaller RF voltage applied to the RF carpet. At the same time, the extraction efficiency of the molecular contaminant drastically drops at voltages below $ 55 $ V$ _{\mathrm{peak-peak}} $. At RF voltages of $ 37 $ V$ _{\mathrm{peak-peak}} $ and $ 41 $ V$ _{\mathrm{peak-peak}} $, the molecular contaminant is suppressed by factors of 220 and 160 stronger compared to the stable Kr isotopes, respectively. The data points for the molecular background reflect the combined efficiency of all other ions measured in the mass-to-charge range from 70 u/e to 95 u/e. The overall background is also strongly suppressed with decreasing RF voltage. Its slope is not as steep as of the molecular contaminant at 80 u/e, because the background is composed of multiple contaminants that all have different ion mobilities. Changes in the steepness of the slope indicate the removal of individual species from the mixture. The different behavior observed for $ ^{78} $Kr suggests that its mass line also overlapped with molecular contaminants, which were not recognized in the measured spectra due to the low statistics and resolving power. For RF voltages lower than $ 48 $ V$ _{\mathrm{peak-peak}} $, the efficiency of $ ^{78} $Kr follows the same trend as for the other Kr isotopes, with an offset due to the removed overlapping contaminants. From the difference in the RF voltage required for $50~\%$ extraction efficiency, it is possible to deduce the difference in the reduced ion mobility of $^{83}$Kr and the molecular contaminant at 80~u/e using Eq.~(\ref{Eq_1_19}). The mobility ratio of $K_0(^{83}\mathrm{Kr})/K_0(\mathrm{MC80})=1.4\pm0.2$ is obtained. The relatively large uncertainty of this value is caused by the uncertainties in the measured counts due to the dead-time effects. According to \cite{Viehl1995}, the $ ^{83} $Kr$^+$ ions have a reduced ion mobility in helium gas at 74 K of about 17~cm$ ^{2} $/(Vs). The estimated reduced mobility value of the molecular contaminant is, therefore, $K_0(\mathrm{MC80})=(12\pm2)$~cm$ ^{2} $/(Vs). This reduced ion mobility value indicates that the molecule might be a hydrocarbon. However, there is no corresponding isotopic abundance pattern seen in any of the measured spectra. The measured abundances of the peak at 81~u/e, which would correspond to the same molecule with one $^{12}$C replaced by $^{13}$C, are less than $2\%$ of the abundances of MC80. A high-resolution measurement of this mass-to-charge range with the MR-TOF-MS was not performed during the discussed experiment, so the identification of this molecular contaminant was not done.

\begin{figure}[]
\centering
\includegraphics[width=0.45\textwidth]{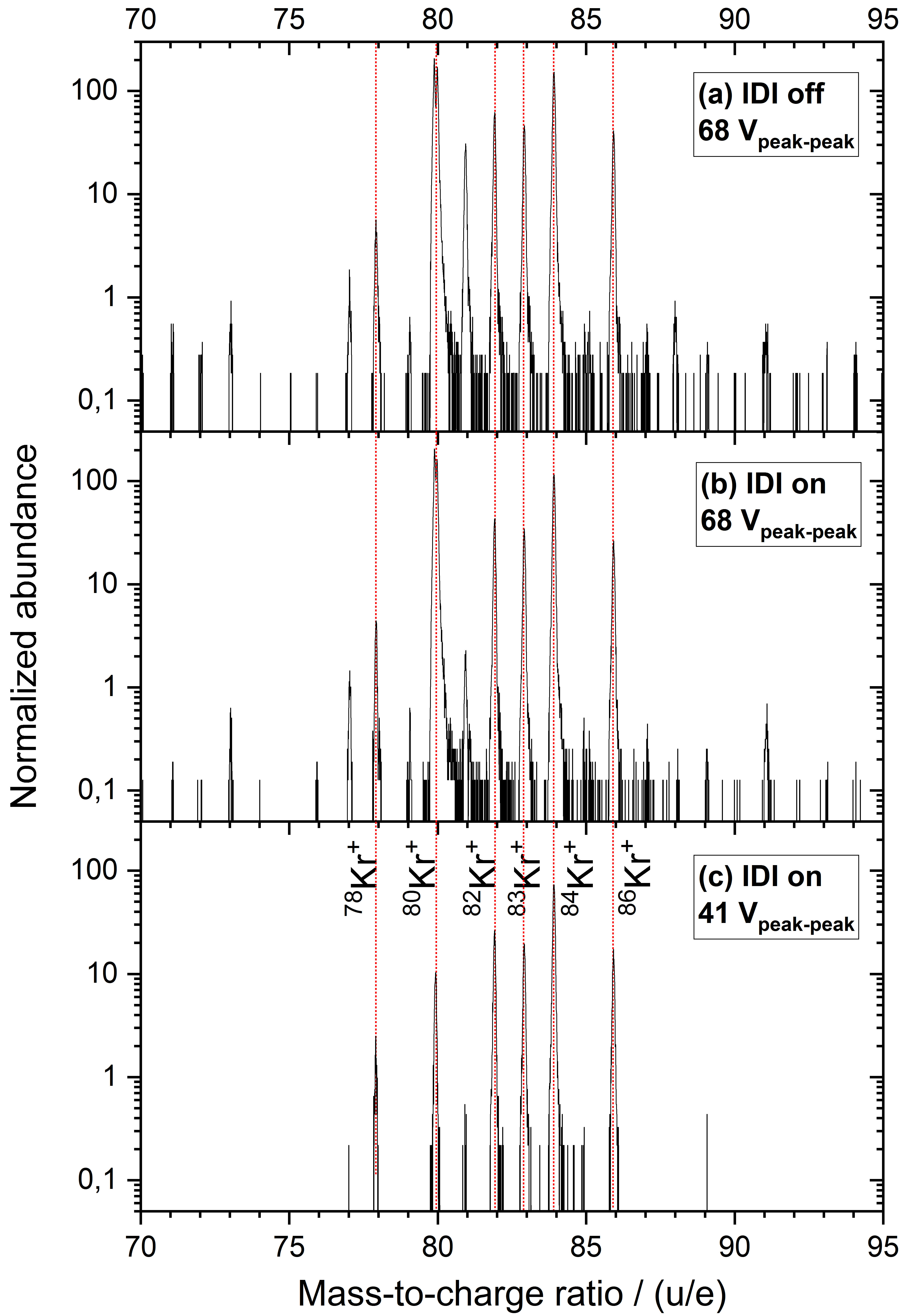}
\caption{Measured mass spectra of ionization products extracted out of the CSC plotted in logarithmic scale. All spectra are normalized to the number of spills and intensity of the incoming ion beam. a) IDI is not applied, RF carpet is at 68~V$ _{\mathrm{peak-peak}} $; b) IDI is applied, RF carpet is at 68~V$ _{\mathrm{peak-peak}} $ (spectrum from Fig.~\ref{Fig_2}a); c) IDI is applied, RF carpet is at 41~V$ _{\mathrm{peak-peak}} $ (spectrum from Fig.~\ref{Fig_2}c). Separation by ion mobility at the RF carpet strongly suppresses the molecular background over a mass-to-charge range of 25~u/e.}
\label{Fig_3_14}
\end{figure}

In Figure~\ref{Fig_3_14}, the spectra for $ 68 $ V$ _{\mathrm{peak-peak}} $ and $ 41 $~V$ _{\mathrm{peak-peak}} $ (Fig.~\ref{Fig_2}a and Fig.~\ref{Fig_2}c) are shown on a logarithmic scale. The logarithmic scale reveals low-abundant molecular contaminants, which could not be seen on the linear scale. The spectrum measured with $ 68 $ V$ _{\mathrm{peak-peak}} $ at the RF carpet and no IDI applied is also added to the plot. As can be seen from the first two panels, in this mass-to-charge range, the IDI technique did not bring significant improvement to the composition of the spectrum. Only the contaminant peak at 81 u/e was suppressed by an order of magnitude. At the same time, it can be clearly seen that the separation by ion mobility at the RF carpet works over a very broad mass-to-charge range. The ratio of counts of Kr isotopes to all other counts measured in the range from 70 u/e to 95 u/e in Fig.~\ref{Fig_3_14}b and Fig.~\ref{Fig_3_14}c increases by a factor of 125. In Figure ~\ref{Fig_3_14}c, background events amount to less than $1~\%$ of all measured counts. Both very abundant and low-abundant molecular contaminants are essentially completely removed from the spectrum. These background-free conditions are of a high importance for the measurements of low-yield nuclides as they improve the overall sensitivity of the experiment.

\section{Conclusion and Outlook}
The separation by ion mobility in gas-filled stopping cells with RF structures was developed and successfully implemented at the FRS-IC. In this work, it was shown for an RF carpet, however, similar results can be also expected for RF funnels. Suppression of molecular contaminants by almost three orders of magnitude for less than $50~\%$ loss of the ions of interest was demonstrated. A background level of less than $1~\%$ for a very broad mass-to-charge range of 25 u/e was achieved. Neutralization of molecular ions at the RF carpet surface helps to reduce unnecessary space charge effects at the extraction nozzle and in the downstream beamline. In addition, it can be used to remove strongly-bound molecules that are difficult to break up by collision-induced dissociation in the RFQ beamline. In combination with other separation methods, this technique can be a very powerful tool for the suppression of molecular background and significant improvement of the cleanliness of the measured spectrum. Although the ion mobility value of the molecular contaminant obtained in this work ($K_0(\mathrm{MC80})=(12\pm2$)~cm$ ^{2} $/(Vs)) has a relatively large uncertainty due to the dead-time effects, the developed technique can also be, in principle, used as a complementary method for the measurement of ion mobilities.

It is important to note that the developed technique requires an extended understanding of the interplay of the system’s operating conditions and parameters. In addition, the current RF carpet is resonance-driven, so the applied RF voltage and, therefore, the separation by ion mobility may be affected by small temperature changes. This will be avoided in the future by implementing an active RF-amplitude and areal density regulation.

\section*{Acknowledgments}
This work was supported by the German Federal Ministry for Education and Research (BMBF) under contracts no.\ 05P16RGFN1 and 05P19RGFN1, by Justus Liebig University Gie{\ss}en and GSI under the JLU-GSI strategic Helmholtz partnership agreement, by HGS-HIRe, and by the Hessian Ministry for Science and Art (HMWK) through the LOEWE Center HICforFAIR. PC is supported by ELI-NP Phase II (1/07.07.2016, COP, ID 1334).


\appendix

\bibliographystyle{elsarticle-num-names}
\bibliography{IONAS_2019-11.tex}

\begin{thebibliography}{34}
\expandafter\ifx\csname natexlab\endcsname\relax\def\natexlab#1{#1}\fi
\providecommand{\url}[1]{\texttt{#1}}
\providecommand{\href}[2]{#2}
\providecommand{\path}[1]{#1}
\providecommand{\DOIprefix}{doi:}
\providecommand{\ArXivprefix}{arXiv:}
\providecommand{\URLprefix}{URL: }
\providecommand{\Pubmedprefix}{pmid:}
\providecommand{\doi}[1]{\href{http://dx.doi.org/#1}{\path{#1}}}
\providecommand{\Pubmed}[1]{\href{pmid:#1}{\path{#1}}}
\providecommand{\bibinfo}[2]{#2}
\ifx\xfnm\relax \def\xfnm[#1]{\unskip,\space#1}\fi
\bibitem[{Wada(2013)}]{Wada2013}
\bibinfo{author}{M.~Wada},
\newblock \bibinfo{title}{Genealogy of gas cells for low-energy {RI}-beam
  production},
\newblock \bibinfo{journal}{Nucl. Instrum. Methods B} \bibinfo{volume}{317}
  (\bibinfo{year}{2013}) \bibinfo{pages}{450--456}.
\bibitem[{Wada et~al.(2003)Wada, Ishida, Nakamura, Yamazaki, Kambara, Ohyama,
  Kanai, Kojima, Nakai, Ohshima, Yoshida, Kubo, Matsuo, Fukuyama, Okada,
  Sonoda, Ohtani, Noda, Kawakami, and Katayama}]{Wada2003}
\bibinfo{author}{M.~Wada}, \bibinfo{author}{Y.~Ishida},
  \bibinfo{author}{T.~Nakamura}, \bibinfo{author}{Y.~Yamazaki},
  \bibinfo{author}{T.~Kambara}, \bibinfo{author}{H.~Ohyama},
  \bibinfo{author}{Y.~Kanai}, \bibinfo{author}{T.~M. Kojima},
  \bibinfo{author}{Y.~Nakai}, \bibinfo{author}{N.~Ohshima},
  \bibinfo{author}{A.~Yoshida}, \bibinfo{author}{T.~Kubo},
  \bibinfo{author}{Y.~Matsuo}, \bibinfo{author}{Y.~Fukuyama},
  \bibinfo{author}{K.~Okada}, \bibinfo{author}{T.~Sonoda},
  \bibinfo{author}{S.~Ohtani}, \bibinfo{author}{K.~Noda},
  \bibinfo{author}{H.~Kawakami}, \bibinfo{author}{I.~Katayama},
\newblock \bibinfo{title}{Slow {RI}-beams from projectile fragment separators},
\newblock \bibinfo{journal}{Nucl. Instrum. Methods B} \bibinfo{volume}{204}
  (\bibinfo{year}{2003}) \bibinfo{pages}{570--581}.
\bibitem[{Savard et~al.(2003)Savard, Clark, Boudreau, Buchinger, Crawford,
  Geissel, Greene, Gulick, Heinz, Lee, Levand, Maier, M\"unzenberg,
  Scheidenberger, Seweryniak, Sharma, Sprouse, Vaz, Wang, Zabransky, and
  Zhou}]{Savar2003}
\bibinfo{author}{G.~Savard}, \bibinfo{author}{J.~Clark},
  \bibinfo{author}{C.~Boudreau}, \bibinfo{author}{F.~Buchinger},
  \bibinfo{author}{J.~Crawford}, \bibinfo{author}{H.~Geissel},
  \bibinfo{author}{J.~Greene}, \bibinfo{author}{S.~Gulick},
  \bibinfo{author}{A.~Heinz}, \bibinfo{author}{J.~K.~P. Lee},
  \bibinfo{author}{A.~Levand}, \bibinfo{author}{M.~Maier},
  \bibinfo{author}{G.~M\"unzenberg}, \bibinfo{author}{C.~Scheidenberger},
  \bibinfo{author}{D.~Seweryniak}, \bibinfo{author}{K.~S. Sharma},
  \bibinfo{author}{G.~Sprouse}, \bibinfo{author}{J.~Vaz},
  \bibinfo{author}{J.~C. Wang}, \bibinfo{author}{B.~J. Zabransky},
  \bibinfo{author}{Z.~Zhou},
\newblock \bibinfo{title}{Development and operation of gas catchers to
  thermalize fusion-evaporation and fragmentation products},
\newblock \bibinfo{journal}{Nucl. Instrum. Methods B} \bibinfo{volume}{204}
  (\bibinfo{year}{2003}) \bibinfo{pages}{582--586}.
\bibitem[{Weissman et~al.(2005)Weissman, Morrissey, Bollen, Davies, Kwan, Lofy,
  Schury, Schwarz, Sumithrarachchi, Sun, and Ringle}]{Weiss2005}
\bibinfo{author}{L.~Weissman}, \bibinfo{author}{D.~J. Morrissey},
  \bibinfo{author}{G.~Bollen}, \bibinfo{author}{D.~A. Davies},
  \bibinfo{author}{E.~Kwan}, \bibinfo{author}{P.~A. Lofy},
  \bibinfo{author}{P.~Schury}, \bibinfo{author}{S.~Schwarz},
  \bibinfo{author}{C.~Sumithrarachchi}, \bibinfo{author}{T.~Sun},
  \bibinfo{author}{R.~Ringle},
\newblock \bibinfo{title}{Conversion of 92 {MeV/u} $^{38}${Ca}/$^{37}${K}
  projectile fragments into thermalized ion beams},
\newblock \bibinfo{journal}{Nucl. Instrum. Methods A} \bibinfo{volume}{540}
  (\bibinfo{year}{2005}) \bibinfo{pages}{245--258}.
\bibitem[{Neumayr et~al.(2006)Neumayr, Beck, Habs, Heinz, Szerypo, Thirolf,
  Varentsov, Voit, Ackermann, Beck, Block, Di, Eliseev, Geissel, Herfurth,
  He{\ss}berger, Hofmann, Kluge, Mukherjee, M\"unzenberg, Petrick, Quint,
  Rahaman, Rauth, Rodriguez, Scheidenberger, Sikler, Wang, Weber, Pla{\ss},
  Breitenfeldt, Chaudhuri, Marx, Schweikhard, Dodonov, Novikov, and
  Suhonen}]{Neuma2006a}
\bibinfo{author}{J.~Neumayr}, \bibinfo{author}{L.~Beck},
  \bibinfo{author}{D.~Habs}, \bibinfo{author}{S.~Heinz},
  \bibinfo{author}{J.~Szerypo}, \bibinfo{author}{P.~Thirolf},
  \bibinfo{author}{V.~Varentsov}, \bibinfo{author}{F.~Voit},
  \bibinfo{author}{D.~Ackermann}, \bibinfo{author}{D.~Beck},
  \bibinfo{author}{M.~Block}, \bibinfo{author}{Z.~Di},
  \bibinfo{author}{S.~Eliseev}, \bibinfo{author}{H.~Geissel},
  \bibinfo{author}{F.~Herfurth}, \bibinfo{author}{F.~He{\ss}berger},
  \bibinfo{author}{S.~Hofmann}, \bibinfo{author}{H.-J. Kluge},
  \bibinfo{author}{M.~Mukherjee}, \bibinfo{author}{G.~M\"unzenberg},
  \bibinfo{author}{M.~Petrick}, \bibinfo{author}{W.~Quint},
  \bibinfo{author}{S.~Rahaman}, \bibinfo{author}{C.~Rauth},
  \bibinfo{author}{D.~Rodriguez}, \bibinfo{author}{C.~Scheidenberger},
  \bibinfo{author}{G.~Sikler}, \bibinfo{author}{Z.~Wang},
  \bibinfo{author}{C.~Weber}, \bibinfo{author}{W.~Pla{\ss}},
  \bibinfo{author}{M.~Breitenfeldt}, \bibinfo{author}{A.~Chaudhuri},
  \bibinfo{author}{G.~Marx}, \bibinfo{author}{L.~Schweikhard},
  \bibinfo{author}{A.~Dodonov}, \bibinfo{author}{Y.~Novikov},
  \bibinfo{author}{M.~Suhonen},
\newblock \bibinfo{title}{The ion-catcher device for {SHIPTRAP}},
\newblock \bibinfo{journal}{Nucl. Instrum. Methods B} \bibinfo{volume}{244}
  (\bibinfo{year}{2006}) \bibinfo{pages}{489--500}.
\bibitem[{Ranjan et~al.(2011)Ranjan, Purushothaman, Dickel, Geissel, Pla{\ss},
  Sch\"afer, Scheidenberger, {Van de Walle}, Weick, and Dendooven}]{Ranja2011}
\bibinfo{author}{M.~Ranjan}, \bibinfo{author}{S.~Purushothaman},
  \bibinfo{author}{T.~Dickel}, \bibinfo{author}{H.~Geissel},
  \bibinfo{author}{W.~R. Pla{\ss}}, \bibinfo{author}{D.~Sch\"afer},
  \bibinfo{author}{C.~Scheidenberger}, \bibinfo{author}{J.~{Van de Walle}},
  \bibinfo{author}{H.~Weick}, \bibinfo{author}{P.~Dendooven},
\newblock \bibinfo{title}{New stopping cell capabilities: {RF} carpet
  performance at high gas density and cryogenic operation},
\newblock \bibinfo{journal}{Eur. Phys. Lett.} \bibinfo{volume}{96}
  (\bibinfo{year}{2011}) \bibinfo{pages}{52001}.
\bibitem[{Kudryavtsev et~al.(2001)Kudryavtsev, Bruyneel, Huyse, Gentens,
  Van~den Bergh, Van~Duppen, and Vermeeren}]{Kudry2001}
\bibinfo{author}{Y.~Kudryavtsev}, \bibinfo{author}{B.~Bruyneel},
  \bibinfo{author}{M.~Huyse}, \bibinfo{author}{J.~Gentens},
  \bibinfo{author}{P.~Van~den Bergh}, \bibinfo{author}{P.~Van~Duppen},
  \bibinfo{author}{L.~Vermeeren},
\newblock \bibinfo{title}{A gas cell for thermalizing, storing and transporting
  radioactive ions and atoms. part i: Off-line studies with a laser ion
  source},
\newblock \bibinfo{journal}{Nucl. Instrum. Methods B} \bibinfo{volume}{179}
  (\bibinfo{year}{2001}) \bibinfo{pages}{412--435}.
\bibitem[{McLuckey(1992)}]{Mcluc1992}
\bibinfo{author}{S.~A. McLuckey},
\newblock \bibinfo{title}{Principles of collisional activation in analytical
  mass spectrometry},
\newblock \bibinfo{journal}{J. Am. Soc. Mass Spectrom.} \bibinfo{volume}{3}
  (\bibinfo{year}{1992}) \bibinfo{pages}{599--614}.
\bibitem[{Schury et~al.(2006)Schury, Bollen, Block, Morrissey, Ringle, Prinke,
  Savory, Schwarz, and Sun}]{Schur2006}
\bibinfo{author}{P.~Schury}, \bibinfo{author}{G.~Bollen},
  \bibinfo{author}{M.~Block}, \bibinfo{author}{D.~Morrissey},
  \bibinfo{author}{R.~Ringle}, \bibinfo{author}{A.~Prinke},
  \bibinfo{author}{J.~Savory}, \bibinfo{author}{S.~Schwarz},
  \bibinfo{author}{T.~Sun},
\newblock \bibinfo{title}{Beam purification techniques for low energy rare
  isotope beams from a gas cell},
\newblock \bibinfo{journal}{Hyperfine Interact.} \bibinfo{volume}{173}
  (\bibinfo{year}{2006}) \bibinfo{pages}{165--170}.
\bibitem[{Morrissey(2007)}]{Morri2007}
\bibinfo{author}{D.~J. Morrissey},
\newblock \bibinfo{title}{Extraction of thermalized projectile fragments from
  gas},
\newblock \bibinfo{journal}{Eur. Phys. J. Special Topics} \bibinfo{volume}{150}
  (\bibinfo{year}{2007}) \bibinfo{pages}{365--366}.
\bibitem[{Moore(2008)}]{Moore2008}
\bibinfo{author}{I.~D. Moore},
\newblock \bibinfo{title}{New concepts for the ion guide technique},
\newblock \bibinfo{journal}{Nucl. Instrum. Methods B} \bibinfo{volume}{266}
  (\bibinfo{year}{2008}) \bibinfo{pages}{4434--4441}.
\bibitem[{Takamine et~al.(2005)Takamine, Wada, Ishida, Nakamura, Okada,
  Yamazaki, Kambara, Kanai, Kojima, Nakai, Oshima, Yoshida, Kubo, Ohtani, Noda,
  Katayama, Hostain, Varentsov, and Wollnik}]{Takam2005}
\bibinfo{author}{A.~Takamine}, \bibinfo{author}{M.~Wada},
  \bibinfo{author}{Y.~Ishida}, \bibinfo{author}{T.~Nakamura},
  \bibinfo{author}{K.~Okada}, \bibinfo{author}{Y.~Yamazaki},
  \bibinfo{author}{T.~Kambara}, \bibinfo{author}{Y.~Kanai},
  \bibinfo{author}{T.~M. Kojima}, \bibinfo{author}{Y.~Nakai},
  \bibinfo{author}{N.~Oshima}, \bibinfo{author}{A.~Yoshida},
  \bibinfo{author}{T.~Kubo}, \bibinfo{author}{S.~Ohtani},
  \bibinfo{author}{K.~Noda}, \bibinfo{author}{I.~Katayama},
  \bibinfo{author}{P.~Hostain}, \bibinfo{author}{V.~Varentsov},
  \bibinfo{author}{H.~Wollnik},
\newblock \bibinfo{title}{Space-charge effects in the catcher gas cell of a
  {RF} ion guide},
\newblock \bibinfo{journal}{Rev. Sci. Instrum.} \bibinfo{volume}{76}
  (\bibinfo{year}{2005}) \bibinfo{pages}{103503}.
\bibitem[{Reiter et~al.(2016)Reiter, Rink, Dickel, Haettner, Hei{\ss}e,
  Pla{\ss}, Purushothaman, Amjad, {Ayet San Andr{\'e}s}, Bergmann, Blum,
  Dendooven, Diwisch, Ebert, Geissel, Greiner, Hornung, Jesch,
  Kalantar-Nayestanaki, Kn\"obel, Lang, Lippert, Miskun, Moore, Nociforo,
  Petrick, Pietri, Pf\"utzner, Pohjalainen, Prochazka, Scheidenberger, Takechi,
  Tanaka, Weick, Winfield, and Xu}]{Reite2016}
\bibinfo{author}{M.~P. Reiter}, \bibinfo{author}{A.-K. Rink},
  \bibinfo{author}{T.~Dickel}, \bibinfo{author}{E.~Haettner},
  \bibinfo{author}{F.~Hei{\ss}e}, \bibinfo{author}{W.~R. Pla{\ss}},
  \bibinfo{author}{S.~Purushothaman}, \bibinfo{author}{F.~Amjad},
  \bibinfo{author}{S.~{Ayet San Andr{\'e}s}}, \bibinfo{author}{J.~Bergmann},
  \bibinfo{author}{D.~Blum}, \bibinfo{author}{P.~Dendooven},
  \bibinfo{author}{M.~Diwisch}, \bibinfo{author}{J.~Ebert},
  \bibinfo{author}{H.~Geissel}, \bibinfo{author}{F.~Greiner},
  \bibinfo{author}{C.~Hornung}, \bibinfo{author}{C.~Jesch},
  \bibinfo{author}{N.~Kalantar-Nayestanaki}, \bibinfo{author}{R.~Kn\"obel},
  \bibinfo{author}{J.~Lang}, \bibinfo{author}{W.~Lippert},
  \bibinfo{author}{I.~Miskun}, \bibinfo{author}{I.~D. Moore},
  \bibinfo{author}{C.~Nociforo}, \bibinfo{author}{M.~Petrick},
  \bibinfo{author}{S.~Pietri}, \bibinfo{author}{M.~Pf\"utzner},
  \bibinfo{author}{I.~Pohjalainen}, \bibinfo{author}{A.~Prochazka},
  \bibinfo{author}{C.~Scheidenberger}, \bibinfo{author}{M.~Takechi},
  \bibinfo{author}{Y.~K. Tanaka}, \bibinfo{author}{H.~Weick},
  \bibinfo{author}{J.~S. Winfield}, \bibinfo{author}{X.~Xu},
\newblock \bibinfo{title}{Rate capability of a cryogenic stopping cell for
  uranium projectile fragments produced at 1000 {MeV/u}},
\newblock \bibinfo{journal}{Nucl. Instrum. Methods B} \bibinfo{volume}{376}
  (\bibinfo{year}{2016}) \bibinfo{pages}{240--245}.
\bibitem[{Sumithrarachchi et~al.(2020)Sumithrarachchi, Morrissey, Schwarz,
  Lund, Bollen, Ringle, Savard, and Villari}]{Sumit2020}
\bibinfo{author}{C.~Sumithrarachchi}, \bibinfo{author}{D.~Morrissey},
  \bibinfo{author}{S.~Schwarz}, \bibinfo{author}{K.~Lund},
  \bibinfo{author}{G.~Bollen}, \bibinfo{author}{R.~Ringle},
  \bibinfo{author}{G.~Savard}, \bibinfo{author}{A.~Villari},
\newblock \bibinfo{title}{Beam thermalization in a large gas catcher},
\newblock \bibinfo{journal}{Nucl. Instrum. Methods B} \bibinfo{volume}{463}
  (\bibinfo{year}{2020}) \bibinfo{pages}{305--309}.
\bibitem[{Eiceman et~al.(2013)Eiceman, Karpas, and Hill~Jr}]{Eicem2013}
\bibinfo{author}{G.~A. Eiceman}, \bibinfo{author}{Z.~Karpas},
  \bibinfo{author}{H.~H. Hill~Jr}, \bibinfo{title}{Ion mobility spectrometry},
  \bibinfo{publisher}{CRC press}, \bibinfo{year}{2013}.
\bibitem[{Kelly et~al.(2010)Kelly, Tolmachev, Page, Tang, and
  Smith}]{Kelly2010}
\bibinfo{author}{R.~T. Kelly}, \bibinfo{author}{A.~V. Tolmachev},
  \bibinfo{author}{J.~S. Page}, \bibinfo{author}{K.~Tang},
  \bibinfo{author}{R.~D. Smith},
\newblock \bibinfo{title}{The ion funnel: theory, implementations, and
  applications},
\newblock \bibinfo{journal}{Mass spectrometry reviews} \bibinfo{volume}{29}
  (\bibinfo{year}{2010}) \bibinfo{pages}{294--312}.
\bibitem[{Tolmachev et~al.(1997)Tolmachev, Chernushevich, Dodonov, and
  Standing}]{Tolma1997}
\bibinfo{author}{A.~V. Tolmachev}, \bibinfo{author}{I.~V. Chernushevich},
  \bibinfo{author}{A.~F. Dodonov}, \bibinfo{author}{K.~G. Standing},
\newblock \bibinfo{title}{A collisional focusing ion guide for coupling an
  atmospheric pressure ion source to a mass spectrometer},
\newblock \bibinfo{journal}{Nucl. Instrum. Methods B} \bibinfo{volume}{124}
  (\bibinfo{year}{1997}) \bibinfo{pages}{112--119}.
\bibitem[{Viehland(2012)}]{Viehl2012}
\bibinfo{author}{L.~A. Viehland},
\newblock \bibinfo{title}{Zero-field mobilities in helium: highly accurate
  values for use in ion mobility spectrometry},
\newblock \bibinfo{journal}{Int. J. Ion Mobil. Spec.} \bibinfo{volume}{15}
  (\bibinfo{year}{2012}) \bibinfo{pages}{21--29}.
\bibitem[{Visentin et~al.(2020)Visentin, Laatiaoui, Viehland, and
  Buchachenko}]{Visen2020}
\bibinfo{author}{G.~Visentin}, \bibinfo{author}{M.~Laatiaoui},
  \bibinfo{author}{L.~A. Viehland}, \bibinfo{author}{A.~A. Buchachenko},
\newblock \bibinfo{title}{Mobility of the singly-charged lanthanide and
  actinide cations: Trends and perspectives},
\newblock \bibinfo{journal}{Frontiers in Chemistry} \bibinfo{volume}{8}
  (\bibinfo{year}{2020}) \bibinfo{pages}{438}.
\bibitem[{Patterson(1970)}]{Patte1970}
\bibinfo{author}{P.~L. Patterson},
\newblock \bibinfo{title}{Temperature dependence of helium-ion mobilities},
\newblock \bibinfo{journal}{Phys. Rev. A} \bibinfo{volume}{2}
  (\bibinfo{year}{1970}) \bibinfo{pages}{1154}.
\bibitem[{Pla{\ss} et~al.(2013)Pla{\ss}, Dickel, Purushothaman, Dendooven,
  Geissel, Ebert, Haettner, Jesch, Ranjan, Reiter, Weick, Amjad, Ayet, Diwisch,
  Estrade, Farinon, Greiner, Kalantar-Nayestanaki, Kn\"obel, Kurcewicz, Lang,
  Moore, Mukha, Nociforo, Petrick, Pfuetzner, Pietri, Prochazka, Rink,
  Rinta-Antila, Sch\"afer, Scheidenberger, Takechi, Tanaka, Winfield, and
  Yavor}]{Plass2013b}
\bibinfo{author}{W.~R. Pla{\ss}}, \bibinfo{author}{T.~Dickel},
  \bibinfo{author}{S.~Purushothaman}, \bibinfo{author}{P.~Dendooven},
  \bibinfo{author}{H.~Geissel}, \bibinfo{author}{J.~Ebert},
  \bibinfo{author}{E.~Haettner}, \bibinfo{author}{C.~Jesch},
  \bibinfo{author}{M.~Ranjan}, \bibinfo{author}{M.~P. Reiter},
  \bibinfo{author}{H.~Weick}, \bibinfo{author}{F.~Amjad},
  \bibinfo{author}{S.~Ayet}, \bibinfo{author}{M.~Diwisch},
  \bibinfo{author}{A.~Estrade}, \bibinfo{author}{F.~Farinon},
  \bibinfo{author}{F.~Greiner}, \bibinfo{author}{N.~Kalantar-Nayestanaki},
  \bibinfo{author}{R.~Kn\"obel}, \bibinfo{author}{J.~Kurcewicz},
  \bibinfo{author}{J.~Lang}, \bibinfo{author}{I.~Moore},
  \bibinfo{author}{I.~Mukha}, \bibinfo{author}{C.~Nociforo},
  \bibinfo{author}{M.~Petrick}, \bibinfo{author}{M.~Pfuetzner},
  \bibinfo{author}{S.~Pietri}, \bibinfo{author}{A.~Prochazka},
  \bibinfo{author}{A.-K. Rink}, \bibinfo{author}{S.~Rinta-Antila},
  \bibinfo{author}{D.~Sch\"afer}, \bibinfo{author}{C.~Scheidenberger},
  \bibinfo{author}{M.~Takechi}, \bibinfo{author}{Y.~K. Tanaka},
  \bibinfo{author}{J.~S. Winfield}, \bibinfo{author}{M.~I. Yavor},
\newblock \bibinfo{title}{The {FRS Ion Catcher} - a facility for high-precision
  experiments with stopped projectile and fission fragments},
\newblock \bibinfo{journal}{Nucl. Instrum. Methods B} \bibinfo{volume}{317}
  (\bibinfo{year}{2013}) \bibinfo{pages}{457--462}.
\bibitem[{Pla{\ss} et~al.(2019)Pla{\ss}, Dickel, Mardor, Pietri, Geissel,
  Scheidenberger, Amanbayev, San~Andr{\'e}s, {\"A}yst{\"o}, Balabanski
  et~al.}]{Plass2019}
\bibinfo{author}{W.~R. Pla{\ss}}, \bibinfo{author}{T.~Dickel},
  \bibinfo{author}{I.~Mardor}, \bibinfo{author}{S.~Pietri},
  \bibinfo{author}{H.~Geissel}, \bibinfo{author}{C.~Scheidenberger},
  \bibinfo{author}{D.~Amanbayev}, \bibinfo{author}{S.~A. San~Andr{\'e}s},
  \bibinfo{author}{J.~{\"A}yst{\"o}}, \bibinfo{author}{D.~L. Balabanski},
  et~al.,
\newblock \bibinfo{title}{The science case of the {FRS} {Ion Catcher} for
  {FAIR} {P}hase-0},
\newblock \bibinfo{journal}{Hyperfine Interactions} \bibinfo{volume}{240}
  (\bibinfo{year}{2019}) \bibinfo{pages}{73}.
\bibitem[{Geissel et~al.(2003)Geissel, Weick, Winkler, M\"unzenberg, Chichkine,
  Yavor, Aumann, Behr, B\"ohmer, Br\"unle, Burkard, Benlliure, Cortina-Gil,
  Chulkov, Dael, Ducret, Emling, Franczak, Friese, Gastineau, Gerl,
  Gernh\"auser, Hellstr\"om, Jonson, Kojouharova, Kulessa, Kindler, Kurz,
  Lommel, Mittig, Moritz, M\"uhle, Nolen, Nyman, Roussell-Chomaz,
  Scheidenberger, Schmidt, Schrieder, Sherrill, Simon, S\"ummerer, Tahir,
  Vysotsky, Wollnik, and Zeller}]{Geiss2003}
\bibinfo{author}{H.~Geissel}, \bibinfo{author}{H.~Weick},
  \bibinfo{author}{M.~Winkler}, \bibinfo{author}{G.~M\"unzenberg},
  \bibinfo{author}{V.~Chichkine}, \bibinfo{author}{M.~Yavor},
  \bibinfo{author}{T.~Aumann}, \bibinfo{author}{K.~H. Behr},
  \bibinfo{author}{M.~B\"ohmer}, \bibinfo{author}{A.~Br\"unle},
  \bibinfo{author}{K.~Burkard}, \bibinfo{author}{J.~Benlliure},
  \bibinfo{author}{D.~Cortina-Gil}, \bibinfo{author}{L.~Chulkov},
  \bibinfo{author}{A.~Dael}, \bibinfo{author}{J.-E. Ducret},
  \bibinfo{author}{H.~Emling}, \bibinfo{author}{B.~Franczak},
  \bibinfo{author}{J.~Friese}, \bibinfo{author}{B.~Gastineau},
  \bibinfo{author}{J.~Gerl}, \bibinfo{author}{R.~Gernh\"auser},
  \bibinfo{author}{M.~Hellstr\"om}, \bibinfo{author}{B.~Jonson},
  \bibinfo{author}{J.~Kojouharova}, \bibinfo{author}{R.~Kulessa},
  \bibinfo{author}{B.~Kindler}, \bibinfo{author}{N.~Kurz},
  \bibinfo{author}{B.~Lommel}, \bibinfo{author}{W.~Mittig},
  \bibinfo{author}{G.~Moritz}, \bibinfo{author}{C.~M\"uhle},
  \bibinfo{author}{J.~A. Nolen}, \bibinfo{author}{G.~Nyman},
  \bibinfo{author}{P.~Roussell-Chomaz}, \bibinfo{author}{C.~Scheidenberger},
  \bibinfo{author}{K.-H. Schmidt}, \bibinfo{author}{G.~Schrieder},
  \bibinfo{author}{B.~Sherrill}, \bibinfo{author}{H.~Simon},
  \bibinfo{author}{K.~S\"ummerer}, \bibinfo{author}{N.~A. Tahir},
  \bibinfo{author}{V.~Vysotsky}, \bibinfo{author}{H.~Wollnik},
  \bibinfo{author}{A.~F. Zeller},
\newblock \bibinfo{title}{The {Super-FRS} project at {GSI}},
\newblock \bibinfo{journal}{Nucl. Instrum. Methods B} \bibinfo{volume}{204}
  (\bibinfo{year}{2003}) \bibinfo{pages}{71--85}.
\bibitem[{Winfield et~al.(2013)Winfield, Geissel, Gerl, M\"unzenberg, Nociforo,
  Pla{\ss}, Scheidenberger, Weick, Winkler, and Yavor}]{Winfi2013}
\bibinfo{author}{J.~S. Winfield}, \bibinfo{author}{H.~Geissel},
  \bibinfo{author}{J.~Gerl}, \bibinfo{author}{G.~M\"unzenberg},
  \bibinfo{author}{C.~Nociforo}, \bibinfo{author}{W.~R. Pla{\ss}},
  \bibinfo{author}{C.~Scheidenberger}, \bibinfo{author}{H.~Weick},
  \bibinfo{author}{M.~Winkler}, \bibinfo{author}{M.~I. Yavor},
\newblock \bibinfo{title}{A versatile high-resolution magnetic spectrometer for
  energy compression, reaction studies and nuclear spectroscopy},
\newblock \bibinfo{journal}{Nucl. Instrum. Methods A} \bibinfo{volume}{704}
  (\bibinfo{year}{2013}) \bibinfo{pages}{76--83}.
\bibitem[{Geissel et~al.(1992)Geissel, Armbruster, Behr, Br\"unle, Burkard,
  Chen, Folger, Franczak, Keller, Klepper, Langenbeck, Nickel, Pfeng,
  Pf\"utzner, Roeckl, Rykaczewski, Schall, Schardt, Scheidenberger, Schmidt,
  Schr\"oter, Schwab, S\"ummerer, Weber, M\"unzenberg, Brohm, Clerc, Fauerbach,
  Gaimard, Grewe, Hanelt, Kn\"odler, Steiner, Voss, Weckenmann, Ziegler, Magel,
  Wollnik, Dufour, Fujita, Vieira, and Sherrill}]{Geiss1992}
\bibinfo{author}{H.~Geissel}, \bibinfo{author}{P.~Armbruster},
  \bibinfo{author}{K.~H. Behr}, \bibinfo{author}{A.~Br\"unle},
  \bibinfo{author}{K.~Burkard}, \bibinfo{author}{M.~Chen},
  \bibinfo{author}{H.~Folger}, \bibinfo{author}{B.~Franczak},
  \bibinfo{author}{H.~Keller}, \bibinfo{author}{O.~Klepper},
  \bibinfo{author}{B.~Langenbeck}, \bibinfo{author}{F.~Nickel},
  \bibinfo{author}{E.~Pfeng}, \bibinfo{author}{M.~Pf\"utzner},
  \bibinfo{author}{E.~Roeckl}, \bibinfo{author}{K.~Rykaczewski},
  \bibinfo{author}{I.~Schall}, \bibinfo{author}{D.~Schardt},
  \bibinfo{author}{C.~Scheidenberger}, \bibinfo{author}{K.~H. Schmidt},
  \bibinfo{author}{A.~Schr\"oter}, \bibinfo{author}{T.~Schwab},
  \bibinfo{author}{K.~S\"ummerer}, \bibinfo{author}{M.~Weber},
  \bibinfo{author}{G.~M\"unzenberg}, \bibinfo{author}{T.~Brohm},
  \bibinfo{author}{H.~G. Clerc}, \bibinfo{author}{M.~Fauerbach},
  \bibinfo{author}{J.~J. Gaimard}, \bibinfo{author}{A.~Grewe},
  \bibinfo{author}{E.~Hanelt}, \bibinfo{author}{B.~Kn\"odler},
  \bibinfo{author}{M.~Steiner}, \bibinfo{author}{B.~Voss},
  \bibinfo{author}{J.~Weckenmann}, \bibinfo{author}{C.~Ziegler},
  \bibinfo{author}{A.~Magel}, \bibinfo{author}{H.~Wollnik},
  \bibinfo{author}{J.~P. Dufour}, \bibinfo{author}{Y.~Fujita},
  \bibinfo{author}{D.~J. Vieira}, \bibinfo{author}{B.~Sherrill},
\newblock \bibinfo{title}{The {GSI} projectile fragment separator ({FRS}): a
  versatile magnetic system for relativistic heavy ions},
\newblock \bibinfo{journal}{Nucl. Instrum. Methods B} \bibinfo{volume}{70}
  (\bibinfo{year}{1992}) \bibinfo{pages}{286 -- 297}.
\bibitem[{Reiter(2015)}]{Reite2015}
\bibinfo{author}{M.~P. Reiter}, \bibinfo{title}{Pilot Experiments with
  Relativistic Uranium Projectile and Fission Fragments Thermalized in a
  Cryogenic Gas-Filled Stopping Cell}, \bibinfo{type}{{PhD} thesis}, Justus
  Liebig University Gie{\ss}en, \bibinfo{year}{2015}.
\bibitem[{Ranjan et~al.(2015)Ranjan, Dendooven, Purushothama, Dickel, Reiter,
  Ayet, Haettner, Moore, Kalantar-Nayestanaki, Geissel, Pla{\ss}, Sch\"afer,
  Scheidenberger, Schreuder, Timersma, {Van de Walle}, and Weick}]{Ranja2015}
\bibinfo{author}{M.~Ranjan}, \bibinfo{author}{P.~Dendooven},
  \bibinfo{author}{S.~Purushothama}, \bibinfo{author}{T.~Dickel},
  \bibinfo{author}{M.~P. Reiter}, \bibinfo{author}{S.~Ayet},
  \bibinfo{author}{E.~Haettner}, \bibinfo{author}{I.~D. Moore},
  \bibinfo{author}{N.~Kalantar-Nayestanaki}, \bibinfo{author}{H.~Geissel},
  \bibinfo{author}{W.~R. Pla{\ss}}, \bibinfo{author}{D.~Sch\"afer},
  \bibinfo{author}{C.~Scheidenberger}, \bibinfo{author}{F.~Schreuder},
  \bibinfo{author}{H.~Timersma}, \bibinfo{author}{J.~{Van de Walle}},
  \bibinfo{author}{H.~Weick},
\newblock \bibinfo{title}{Design, construction and cooling system performance
  of a prototype cryogenic stopping cell for the {Super-FRS} at {FAIR}},
\newblock \bibinfo{journal}{Nucl. Instrum. Methods A} \bibinfo{volume}{770}
  (\bibinfo{year}{2015}) \bibinfo{pages}{87--97}.
\bibitem[{Pla{\ss} et~al.(2008)Pla{\ss}, Dickel, Czok, Geissel, Petrick,
  Reinheimer, Scheidenberger, and Yavor}]{Plass2008}
\bibinfo{author}{W.~R. Pla{\ss}}, \bibinfo{author}{T.~Dickel},
  \bibinfo{author}{U.~Czok}, \bibinfo{author}{H.~Geissel},
  \bibinfo{author}{M.~Petrick}, \bibinfo{author}{K.~Reinheimer},
  \bibinfo{author}{C.~Scheidenberger}, \bibinfo{author}{M.~I. Yavor},
\newblock \bibinfo{title}{Isobar separation by time-of-flight mass spectrometry
  for low-energy radioactive ion beam facilities},
\newblock \bibinfo{journal}{Nucl. Instrum. Methods B} \bibinfo{volume}{266}
  (\bibinfo{year}{2008}) \bibinfo{pages}{4560--4564}.
\bibitem[{Dickel et~al.(2015)Dickel, Pla{\ss}, Becker, Czok, Geissel, Haettner,
  Jesch, Kinsel, Petrick, Scheidenberger, and Yavor}]{Dicke2015a}
\bibinfo{author}{T.~Dickel}, \bibinfo{author}{W.~R. Pla{\ss}},
  \bibinfo{author}{A.~Becker}, \bibinfo{author}{U.~Czok},
  \bibinfo{author}{H.~Geissel}, \bibinfo{author}{E.~Haettner},
  \bibinfo{author}{C.~Jesch}, \bibinfo{author}{W.~Kinsel},
  \bibinfo{author}{M.~Petrick}, \bibinfo{author}{C.~Scheidenberger},
  \bibinfo{author}{M.~I. Yavor},
\newblock \bibinfo{title}{A high-performance multiple-reflection time-of-flight
  mass spectrometer and isobar separator for the research with exotic nuclei},
\newblock \bibinfo{journal}{Nucl. Instrum. Methods A} \bibinfo{volume}{777}
  (\bibinfo{year}{2015}) \bibinfo{pages}{172--188}.
\bibitem[{Miskun et~al.(2019)Miskun, Dickel, Mardor, Hornung, Amanbayev, Ayet
  San~Andr{\'e}s, Bergmann, Ebert, Geissel, G{\'o}rska, Greiner, Haettner,
  Pla{\ss}, Purushothaman, Scheidenberger, Rink, Weick, Bagchi, Constantin,
  Kaur, Lippert, Mei, Moore, Otto, Pietri, Pohjalainen, Prochazka, Rappold,
  Reiter, Tanaka, and Winfield}]{Misku2019a}
\bibinfo{author}{I.~Miskun}, \bibinfo{author}{T.~Dickel},
  \bibinfo{author}{I.~Mardor}, \bibinfo{author}{C.~Hornung},
  \bibinfo{author}{D.~Amanbayev}, \bibinfo{author}{S.~Ayet San~Andr{\'e}s},
  \bibinfo{author}{J.~Bergmann}, \bibinfo{author}{J.~Ebert},
  \bibinfo{author}{H.~Geissel}, \bibinfo{author}{M.~G{\'o}rska},
  \bibinfo{author}{F.~Greiner}, \bibinfo{author}{E.~Haettner},
  \bibinfo{author}{W.~R. Pla{\ss}}, \bibinfo{author}{S.~Purushothaman},
  \bibinfo{author}{C.~Scheidenberger}, \bibinfo{author}{A.-K. Rink},
  \bibinfo{author}{H.~Weick}, \bibinfo{author}{S.~Bagchi},
  \bibinfo{author}{P.~Constantin}, \bibinfo{author}{S.~Kaur},
  \bibinfo{author}{W.~Lippert}, \bibinfo{author}{B.~Mei},
  \bibinfo{author}{I.~Moore}, \bibinfo{author}{J.-H. Otto},
  \bibinfo{author}{S.~Pietri}, \bibinfo{author}{I.~Pohjalainen},
  \bibinfo{author}{A.~Prochazka}, \bibinfo{author}{C.~Rappold},
  \bibinfo{author}{M.~P. Reiter}, \bibinfo{author}{Y.~K. Tanaka},
  \bibinfo{author}{J.~S. Winfield},
\newblock \bibinfo{title}{A novel method for the measurement of half-lives and
  decay branching ratios of exotic nuclei},
\newblock \bibinfo{journal}{Eur. Phys. J. A} \bibinfo{volume}{55}
  (\bibinfo{year}{2019}) \bibinfo{pages}{148}. \URLprefix
  \url{https://doi.org/10.1140/epja/i2019-12837-8}.
  \DOIprefix\doi{10.1140/epja/i2019-12837-8}.
\bibitem[{Miskun(2019)}]{Misku2019b}
\bibinfo{author}{I.~Miskun}, \bibinfo{title}{{A Novel Method for the
  Measurement of Half-Lives and Decay Branching Ratios of Exotic Nuclei with
  the FRS Ion Catcher}}, \bibinfo{type}{{PhD} thesis}, Justus Liebig University
  Gie{\ss}en, \bibinfo{year}{2019}.
\bibitem[{Greiner et~al.(2020)Greiner, Dickel, Ayet San~Andr{\'e}s, Bergmann,
  Constantin, Ebert, Geissel, Haettner, Hornung, Miskun, Lippert, Mardor,
  Moore, Plaß, Purushothaman, Rink, Reiter, Scheidenberger, and
  Weick}]{Grein2019}
\bibinfo{author}{F.~Greiner}, \bibinfo{author}{T.~Dickel},
  \bibinfo{author}{S.~Ayet San~Andr{\'e}s}, \bibinfo{author}{J.~Bergmann},
  \bibinfo{author}{P.~Constantin}, \bibinfo{author}{J.~Ebert},
  \bibinfo{author}{H.~Geissel}, \bibinfo{author}{E.~Haettner},
  \bibinfo{author}{C.~Hornung}, \bibinfo{author}{I.~Miskun},
  \bibinfo{author}{W.~Lippert}, \bibinfo{author}{I.~Mardor},
  \bibinfo{author}{I.~Moore}, \bibinfo{author}{W.~R. Plaß},
  \bibinfo{author}{S.~Purushothaman}, \bibinfo{author}{A.-K. Rink},
  \bibinfo{author}{M.~P. Reiter}, \bibinfo{author}{C.~Scheidenberger},
  \bibinfo{author}{H.~Weick},
\newblock \bibinfo{title}{Removal of molecular contamination in low-energy
  {RIBs} by the isolation-dissociation-isolation method},
\newblock \bibinfo{journal}{Nucl. Instrum. Methods B} \bibinfo{volume}{463}
  (\bibinfo{year}{2020}) \bibinfo{pages}{324--326}.
\bibitem[{Weick et~al.(2018)Weick, Geissel, Iwasa, Scheidenberger, and
  Sanchez}]{Weick2018}
\bibinfo{author}{H.~Weick}, \bibinfo{author}{H.~Geissel},
  \bibinfo{author}{N.~Iwasa}, \bibinfo{author}{C.~Scheidenberger},
  \bibinfo{author}{J.~R. Sanchez},
\newblock \bibinfo{title}{{Improved accuracy of the code ATIMA for energy loss
  of heavy ions in matter}},
\newblock \bibinfo{journal}{GSI Sci. Rep. 2017} \bibinfo{volume}{2018-1}
  (\bibinfo{year}{2018}) \bibinfo{pages}{130 p.}
\bibitem[{Viehland and Mason(1995)}]{Viehl1995}
\bibinfo{author}{L.~A. Viehland}, \bibinfo{author}{E.~A. Mason},
\newblock \bibinfo{title}{Transport properties of gaseous ions over a wide
  energy range, {IV}},
\newblock \bibinfo{journal}{At. Data Nucl. Data Tables} \bibinfo{volume}{60}
  (\bibinfo{year}{1995}) \bibinfo{pages}{37--95}.

\end{thebibliography}





\end{document}